\documentclass[a4paper,11pt]{article}
\usepackage{jcappub} 
\usepackage{lineno}
\usepackage{bbm}
\usepackage{physics}
\usepackage{booktabs}
\usepackage{enumitem}
\usepackage{soul}

\newcommand{\Neff}{\ensuremath{N_{\rm eff}}}
\newcommand{\nue}{\ensuremath{\nu_{\rm e}}}
\newcommand{\numu}{\ensuremath{\nu_{\mu}}}
\newcommand{\nutau}{\ensuremath{\nu_{\tau}}}

\newcommand{\Yp}{\ensuremath{Y_{\rm{P}}}}
\newcommand{\prym}{\texttt{PRyMordial}}
\newcommand{\parth}{\texttt{PArthENoPE}}
\newcommand{\primat}{\texttt{PRIMAT}}
\newcommand{\fortepiano}{\texttt{FortEPiaNO}}
\newcommand{\nudec}{\texttt{NUDEC\_BSM}}

\usepackage{xcolor}
\definecolor{mycolor}{RGB}{191, 3, 244}

\newcommand{\as}[1]{{\color{mycolor} #1}}

\title{\boldmath Big Bang Nucleosynthesis as a probe of non-standard neutrino interactions and non-unitary three-neutrino mixing}

\author[a,b]{Gabriela Barenboim,}
\author[c,d]{Stefano Gariazzo,}
\author[a,b]{and Alberto Sánchez-Vargas}

\affiliation[a]{Instituto de F\'{i}sica Corpuscular, CSIC-Universitat de Val\`{e}ncia, Paterna 46980, Spain}
\affiliation[b]{Departament de F\'{i}sica Te\`{o}rica, Universitat de Val\`{e}ncia, Burjassot 46100, Spain}
\affiliation[c]{Department of Physics, University of Turin, via P.\ Giuria 1, 10125 Turin (TO), Italy \looseness=-1}
\affiliation[d]{Istituto Nazionale di Fisica Nucleare (INFN), Sezione di Torino, via P.\ Giuria 1, 10125 Turin (TO), Italy}
\emailAdd{gabriela.barenboim@uv.es, stefano.gariazzo@unito.it, alberto.sanchez-vargas@uv.es }

\abstract{
In this work we investigate the impact of two phenomenological Beyond the Standard Model (BSM) scenarios concerning the role of neutrinos in the early universe: non-standard neutrino interactions (NSI) and non-unitary three-neutrino mixing. We evaluate the impact of these frameworks on two key cosmological observables: the effective number of relativistic neutrino species (\Neff), related to neutrino decoupling, and the abundances of light elements produced at Big Bang Nucleosynthesis (BBN). 

For the first time, neutrino CC-NSI with quarks and non-unitary three-neutrino mixing are studied in the context of BBN, and the constraints on such interactions are found to be remarkably  restrictive. In particular, the BBN limits are competitive with the ones derived from terrestrial experiments for the non-diagonal CC-NSI parameter $\varepsilon^{udV}_{e \alpha}$, with $\alpha \neq e$ and for the non-unitarity parameter $\alpha_{22}$. In the case of non-unitarity, the combination between neutrino decoupling and BBN imposes stringent constraints that can either mildly favour the existence of New Physics (NP), or reinforce the SM, depending on the choice of the experimental nuclear rates involved in the BBN calculation. These results stress the already noted need for further nuclear rates measurements in order to obtain more robust BBN theoretical predictions.
}

\begin{document}
\maketitle
\flushbottom

\section{Introduction}
\label{sec:intro}
The standard cosmological framework, integrating the Standard Model (SM)
of particle physics, provides a consistent picture of the universe, from the first second after the Big Bang right up to the present day. It explains successfully the observed expansion of the universe, the formation of the Cosmic Microwave Background (CMB) and the measured amounts of the lightest elements -hydrogen, helium, and traces of lithium-, produced in the Big Bang Nucleosynthesis (BBN)~\cite{Alpher:1948ve}.

Our theoretical understanding of the universe provides also highly precise predictions for the decoupling of neutrinos from the primordial plasma, and thus for the effective number of relativistic neutrino species (\Neff). State-of-the-art calculations that include neutrino oscillations and next-to-leading order (NLO) Quantum Electrodynamics (QED) corrections to the primordial plasma, refine this prediction to be $N^{\rm SM}_{\rm eff} = 3.044$~\cite{Bennett:2020zkv, Akita:2020szl, Froustey:2020mcq}. Advances in observational technology have driven the possibility to contrast theory with observations at an unparalleled level of precision, challenging the standard cosmological framework. The most recent Planck~\cite{Aghanim2020planck} data (temperature and polarisation maps and the distortions to the CMB spectrum due to gravitational lensing), in combination with Baryon Acoustic Oscillations (BAO) measurements observe $\Neff = 2.99 ^{+0.34}_{-0.33}$ (95\% CL), consistent with the standard prediction. Current precision is expected to be significantly enhanced by future experiments, aiming to achieve $\sigma(\Neff) \simeq 0.02 - 0.03$ in the case of CMB-S4~\cite{Abazajian2016CMB} or $\sigma(\Neff) \simeq 0.05 - 0.07$ in the case of the Simons Observatory~\cite{SimonsObservatory:2018koc}. 

In turn, the standard Big Bang Nucleosynthesis (SBBN) has become a parameter-free theory: the relic nuclear abundances are just a prediction of the cosmological model. The deuterium abundance observed in distant quasar absorption systems and the helium-4 abundance observed in low-metallicity H II regions show a general agreement with such prediction~\cite{ParticleDataGroup2022}.

Nevertheless, the concordance picture in cosmology presents an intriguing paradox, given the known limitations of the SM. One of its shortcomings is the inability to explain the origin of neutrino masses. To resolve this issue, it is necessary to consider Beyond the Standard Model (BSM) theories, that invoke hypothetical mechanisms to provide masses to neutrinos~\cite{Farzan2017, Dev2019, Escrihuela:2015wra}.

The recent observational and theoretical advancements that have prompted the advent of precision cosmology are challenging the validity of the concordance model. Any deviations from the standard predictions could provide compelling evidence for the existence of New Physics (NP).
Indeed, emerging mild tensions between the predictions of BBN and later CMB observations may already be hinting at underlying inconsistencies~\cite{Pitrou:2020etk, Pitrou2021ResolvingConclusions}. Furthermore, the observed abundance of lithium-7 in metal-poor stars is significantly lower than what BBN predicts~\cite{Sbordone2010Lithium}, a long-standing discrepancy known as the ``lithium problem". However, it remains unclear whether its solution lies in NP or in astrophysical processes affecting lithium depletion in stars~\cite{ParticleDataGroup2022}.
Conversely, if ongoing and future precision cosmological measurements continue to support the established paradigm, this would further strengthen the role of cosmological observables in placing stringent constraints on BSM theories. Such constraints would serve to complement and enhance the searches for NP conducted in terrestrial experiments, collectively advancing our understanding of Nature~\cite{Pitrou2018PrecisionHe4predictions}.

After all, the consistency between the expected value of \Neff, BBN predictions and observational data highlights the robustness of our current understanding, while also providing a fertile ground for exploring BSM scenarios. The interplay between cosmological observations and particle physics underscores the importance of precision cosmology as a tool for probing the fundamental nature of the universe.

The aim of this work is to exploit the potential of cosmology as a probe for BSM scenarios. To this end, we adopted two simple and publicly available numerical codes, \nudec~\cite{Escudero2019, Escudero2020} and \prym~\cite{Burns:2023sgx}, which we modified in order to accommodate NP, with the purpose to examine their implications in the early stages of the universe and their consequences on cosmological observables. This knowledge will then be applied to provide cosmological constraints on BSM theories.

This work is organized as follows. Section~\ref{sec:The Early Universe} reviews the physical phenomena that occur during the initial stages of the universe, emphasising on the neutrino decoupling and BBN aspects that will be affected by NP. In Section~\ref{sec:NP}, two phenomenological BSM frameworks related to the generation of neutrino masses are presented: non-standard neutrino interactions (NSI) and non-unitary three-neutrino mixing. Section~\ref{sec:Results} investigates the impact of these BSM models on BBN, where, for the first time, the effects of neutrino CC-NSI with quarks and non-unitarity are considered. Finally, Section~\ref{sec:Conclusions} presents the conclusions of this work.

From now on, natural units will be adopted ($\hbar = c = k_B = 1$).

\section{The Early Universe}
\label{sec:The Early Universe}
\subsection{Neutrino decoupling}
\label{sec:Neutrino decoupling}

One second after the Big Bang, the universe can be described by a hot, dense plasma, an admixture of interacting particles, both relativistic and non-relativistic. Efficient interactions between particles maintain them in local thermodynamic equilibrium (LTE), and the eventual deviations from it are governed by the Boltzmann equation, 
\begin{equation}
    L[f] \equiv \frac{\partial f}{\partial t} - H p \frac{\partial f}{\partial p} = \mathcal{C}[f],
\label{eq:Boltzmann}
\end{equation}
where $f$ is the distribution function of a given species, $p$ its momentum, $t$ is the cosmic time and $H$ the Hubble rate. The collision term, $\mathcal{C}[f]$, encodes the information of particle scatterings, annihilations or decays. For a specific particle $\psi$, it is defined as~\cite{Dolgov:2002wy}
\begin{multline}
\label{eq:collision term}
    \mathcal{C}[f_\psi] \equiv - \frac{1}{2E_\psi} \sum_{X,Y} \int \prod_i d\Pi_{X_i} \prod_j d\Pi_{Y_j} (2\pi)^4 \delta^4(p_\psi + p_X - p_Y) \\
    \left[ \abs{\mathcal{M}}^2_{\psi+X \rightarrow Y} f_\psi \prod_i f_{X_i} \prod_j (1 \pm f_{Y_j}) - \abs{\mathcal{M}}^2_{Y \rightarrow \psi+X} \prod_j f_{Y_j} (1 \pm f_\psi) \prod_i (1 \pm f_{X_i}) \right],
\end{multline}
where $X$ and $Y$ are generic multi-particle states that interact with $\psi$, $d\Pi_{X_i} = \frac{g_{X_i}}{2E}\frac{d^3p}{(2\pi)^3}$ is the Lorentz-invariant phase space element, $g_i$ is the internal degrees of freedom of the species and $\mathcal{M}$ is the scattering amplitude for each interaction. The sign in $(1 \pm f_i)$ depends on the fermionic ($-$) or bosonic ($+$) nature of the particle.

Neutrinos decouple from the electromagnetic plasma when the expansion rate is so fast that  weak interactions cannot keep equilibrium, at $T_{\nu D} \sim 2$~MeV~\cite{Lesgourgues2013neutrinocosmology} if decoupling were an instantaneous process. Since this is only approximate, neutrinos are not completely decoupled when electron-positron annihilation occurs, and the neutrino spectra is non-thermally distorted. The standard picture of neutrino decoupling depends heavily on all the details of the microphysical description contained in the SM. If unknown relativistic particles, BSM interactions, or other NP in the neutrino sector were active at that time, they may alter this scenario. Their possible impact on the radiation content of the universe is parameterized by the effective number of relativistic neutrino species, \Neff,
\begin{equation}
    \rho_R = \rho_\gamma \left( 1 + \frac{7}{8} \left( \frac{4}{11}\right)^{4/3} N_{\rm eff}\right),
\end{equation}
which is defined as a measure of the excess of radiation energy density ($\rho_R$) compared to the photon energy density ($\rho_\gamma$). Assuming that all three neutrino flavours can be described by the same global temperature $T_\nu$, it is possible to write \Neff~as
\begin{equation}
    N_{\rm eff} = 3 \left( \frac{11}{4}\right)^{4/3} \left( \frac{T_{\nu}}{ T_{\gamma}}\right)^4 .
\label{eq:Neff_temperatures}
\end{equation}
Nevertheless, 
for a full treatment of neutrino oscillations and interactions,
the density matrix formalism~\cite{Sigl:1992fn} is required.
It has been applied in several cases in order to obtain precise estimations of \Neff~\cite{Mangano:2005cc, deSalas:2016ztq,Froustey:2020mcq,Akita:2020szl,Bennett:2020zkv}, at the cost of a large computational effort.

In contrast, the approach followed in Refs.~\cite{Escudero2019, Escudero2020} entails capturing the relevant physics of the early universe, assuming certain approximations that allow a significantly faster \Neff~computation. The publicly available code \nudec\footnote{\url{https://github.com/MiguelEA/nudec_BSM}}, provides a precise computation of the order of $\mathcal{O}(10~\rm s)$ avoiding the density matrix formalism. This approach does not only alleviate the required computational effort but also improves the flexibility of the code to facilitate the incorporation of BSM physics. The approximations that greatly simplify the resolution of the Boltzmann equation are to assume exact Fermi-Dirac (FD) distributions for neutrinos and not to account for neutrino oscillations. Then, the evolution of a species can be described in terms of its temperature, leaving only two to four differential equations to solve, one for the photon temperature, $T_\gamma$, and the others for the neutrino temperatures, $T_{\nu_\alpha}$ (depending on whether different temperatures are considered for \nue, \numu\ and \nutau):
\begin{equation}
    \frac{d T_\gamma}{d t} = - \frac{ 4H\rho_\gamma + 3H(\rho_e + p_e) + \frac{\delta \rho_{\nue} }{\delta t} + 2 \frac{\delta \rho_{\numu}}{\delta t}} { \frac{\partial \rho_\gamma}{\partial T_\gamma} + \frac{\partial \rho_e}{\partial T_\gamma} } ,
\label{eq:dT_gamma}
\end{equation}
\begin{equation}
    \frac{d T_{\nu_\alpha}}{d t} = - HT_{\nu_\alpha} +  \frac{\delta \rho_{\nu_\alpha}}{\delta t} \left/  \frac{\partial\rho_{\nu_\alpha}}{\partial T_{\nu_\alpha}} \right. . 
\label{eq:dT_nu}
\end{equation}
Since electrons are tightly coupled to photons, the entire electromagnetic sector can be described by the first equation. The second can be applied to each neutrino flavour separately or to the whole neutrino fluid, defining a global temperature for neutrinos, $T_{\nu} = T_{\nue} = T_{\numu} = T_{\nutau}$. For the purposes of standard neutrino decoupling, setting a common temperature for \numu\ and \nutau, different from the one of \nue, is the option that best mimics the effect of neutrino oscillations~\cite{Escudero2019, Escudero2020}, by simulating the fact that \nue\ interactions are different and its momentum distribution may be slightly higher than the one of the other neutrinos. Such approach guarantees a value of \Neff~which differs by the ones obtained with the full calculation by less than 0.001~\cite{Mangano:2005cc, deSalas:2016ztq, Bennett:2020zkv}. 

The energy transfer rates $\delta  \rho_{\nu_\alpha} / \delta t = \int g_i E_i \frac{d^3 p_i}{(2 \pi)^3} \mathcal{C}[f_i]$ are a measure of the neutrino energy exchanged by interactions with the rest of the plasma, encoding pair production and annihilation, electron-neutrino and neutrino-neutrino scattering. The transfer rates are proportional to the weak couplings in the SM, which for each flavour ($\alpha=e, \mu, \tau$) are:
\begin{equation}
\begin{array}{ccc}
    g^{\rm SM}_{eL} = \sin^2 \theta_W + 1/2,  \quad g^{\rm SM}_{{(\mu, \tau)} L} = \sin^2 \theta_W - 1/2,  \quad g^{\rm SM}_{\alpha R} = \sin^2 \theta_W,
\end{array}
\label{eq:couplings}
\end{equation}
where the weak-mixing angle, $\theta_W$, satisfies $\sin^2 \theta_W \approx 0.231$. These couplings must be modified in the presence of non-standard neutrino interactions, which may alter the strength of the weak interactions, as it will be seen in Section~\ref{sec:NP}. 

The scattering amplitudes of the interactions are initially integrated analytically, assuming Maxwell-Boltzmann (MB) distributions and massless $e^{\pm}$, and later numerically corrected to include quantum statistics and a non-zero electron mass.
When evolving the electromagnetic plasma, finite temperature (FT) Quantum Electrodynamics (QED) effects at leading order (LO) and next-to-leading order (NLO) are also taken into account, modifying eq.~\eqref{eq:dT_gamma} and thus the evolution of $T_\gamma$.

By solving the differential equations above, we obtain the thermodynamic history of the universe. Given this, \Neff~is simply given by eq.~\eqref{eq:Neff_temperatures} if neutrinos share a common temperature, or by the following equation in the more general case of three different temperatures:
\begin{equation}
        N_{\rm eff} =  \left( \frac{11}  {4}\right)^{4/3} \sum_\alpha \left( \frac{T_{\nu_\alpha}}{ T_{\gamma}}\right)^4 .
\end{equation}

When the effect of the neutrino FD distribution, finite electron mass and FT-QED corrections are included, the fast neutrino decoupling computation adopted here ensures a theoretical precision of $\sim 0.001$ on \Neff, corresponding to a relative uncertainty of less than 0.1\% ~\cite{Escudero2020}. Such precision is better than the estimated sensitivity of future cosmological probes \cite{Abazajian2016CMB,SimonsObservatory:2018koc}.

\subsection{Big Bang Nucleosynthesis}
\label{sec:BBN}


At temperatures $T \gtrsim$~MeV, nucleons are essentially the only baryons that exist and charged current weak interactions efficiently convert neutrons into protons and vice versa, maintaining them in chemical equilibrium through the following interactions:

\begin{subequations}
\label{eq:weak interactions}
\vspace{-0.5cm}
\begin{align}
n + \nu_e &\leftrightarrow p + e^- \label{eq:neutron+nu} \\
n &\leftrightarrow p + e^- + \bar{\nu}_e  \label{eq:neutron decay} \\
n + e^+ &\leftrightarrow p + \bar{\nu}_e \label{eq:neutron+positron}
\end{align}
\end{subequations}
The neutron-to-proton ratio, $n_n/n_p$, strongly influences the production of light elements, in particular helium-4, given that mostly all nucleons end up forming it. When the equilibrium forcing $n \leftrightarrow p$ falls out, it is necessary to apply the following Boltzmann equation to study the evolution of nucleon abundances: 
\begin{subequations}
\label{eq:np conversion}
\vspace*{-0.25cm}
\begin{align}
\dot{n}_n + 3H n_n &= -n_n \Gamma_{n \rightarrow p} + n_p \Gamma_{p \rightarrow n} \label{eq:num_density_n} ,\\ \dot{n}_p + 3H n_p &= -n_p \Gamma_{p \rightarrow n} + n_n \Gamma_{n \rightarrow p}\label{eq:num_density_p} ,
\end{align}
\end{subequations}
where the weak rates $\Gamma_{a \rightarrow b}$ capture the physics of $n \leftrightarrow p$ conversion and are proportional to a normalisation factor~\cite{Pitrou2018PrecisionHe4predictions},
\begin{equation}
\label{eq:K_theoretical}
    K \equiv \frac{4 G^2_F V_{ud}^2}{(2\pi)^3} (1 + 3 g^2_A) .
\end{equation}
$G_F$ is the Fermi constant and $V_{ud}$ is the element of the mixing matrix in the quark sector, the Cabbibo-Kobayashi-Maskawa (CKM) matrix, which relates the up and down quarks. The axial current constant for the nucleons, $g_A$, is the primary source of uncertainty in $K$. An alternative and more precise method (by a factor three~\cite{Pitrou2018PrecisionHe4predictions}) for estimating $K$ is through the experimental measurement of the neutron lifetime, $\tau^{\rm exp}_n$. Adopting this approach implies that $\tau^{\rm exp}_n$ already incorporates the potential effects of BSM scenarios that alter the weak rates~\cite{ManganoNSI}. This approach renders $K$ insensitive to the presence of such NP, which is convenient if they are not the subject of study. Conversely, we will follow  eq.~\eqref{eq:K_theoretical} 
for the specific purpose of investigating said BSM scenarios in the early universe. Note that both approaches must address the tensions in the experimental determinations of the constants, either in $\tau^{\rm exp}_n$ (``the neutron lifetime puzzle'')~\cite{Chowdhury:2022ahn} or in $V_{ud}$ (``the Cabibbo angle anomaly'')~\cite{Cirigliano:2022yyo}.

Until the universe cools down to temperatures close to $T_{\rm BBN} \sim 0.1$~MeV, nuclear species heavier than neutrons or protons are in nuclear statistical equilibrium (NSE), and represent an insignificant fraction of baryons. 
The arrest of the deuterium bottleneck marks the ignition of a series of nuclear processes that eventually lead to the primordial abundances of light elements. They are quantified in terms of their number density relative to that of baryons, $X_i  \equiv n_i / n_B$ where $i$ = $^4$He, D, $^3$He, $^7$Li, are the most produced nuclear species in BBN. They are typically normalised to the hydrogen abundance, i.e, to protons ($i/\rm{H} \equiv X_i/X_p$), except for the helium-4 abundance, for which it is customary to define the (approximate) helium-mass fraction, $\Yp \equiv 4X_{^4\rm{He}}$~\cite{Dodelson:2003ft}. At $T_{\rm BBN}$, roughly all free neutrons are locked into helium-4, and the remainder are converted into traces of deuterium and helium-3, and lithium-7 in smaller quantities.
When one computes the Boltzmann equation for nuclides, one obtains a set of $N_{\rm nuc}$ differential equations, which are solved up to the $\mathcal{O}$(keV) era.
For two-body reactions, such equations read as follows~\cite{Iocco:2008va}:
\begin{equation}
\label{eq:dot_X_i}
    \dot{X}_i = \sum_{j,k,l} N_i \left( \Gamma_{kl \rightarrow ij} \frac{X_k^{N_k} X_l^{N_l}}{N_k! N_l!} - \Gamma_{ij \rightarrow kl} \frac{X_i^{N_i} X_j^{N_j}}{N_i! N_j!} \right) \equiv \Gamma_i ,
\end{equation}
with $i, j, k, l$ representing the nuclear species, $N_i$ denotes the stoichiometric coefficient of the nuclide in a given reaction and the nuclear reaction rates are symbolized by $\Gamma_i$. Note that when applied to nucleons, eq.~\eqref{eq:np conversion} is recovered. In the case of a typical BBN reaction, $i + j \rightarrow k + l$, all stoichiometric coefficients are equal to one, and the rate is simply given by $\Gamma_{i + j \rightarrow k + l} = \langle \sigma_{i + j \rightarrow k + l} v \rangle$, where $\sigma$ is the thermally averaged cross-section and multiplies the $i - j$ relative velocity. In practice, laboratory experiments are able to probe the energy range of BBN and provide the data for the nuclear rates, while the reverse reaction rates can be obtained from a detailed balance condition. The nuclear input is capital since the uncertainty of the BBN theoretical predictions is closely tied to the precision of the measurements of the nuclear rates. 

The nuclear network that leads to BBN comprises 424 reactions in total, although the majority can be disregarded due to their negligible impact on the BBN outcome. 
The formation of helium-4 is fundamentally determined by the neutron-to-proton ratio at the onset of BBN, rendering it particularly insensitive to the details of said nuclear network. However, the theoretical uncertainty of its abundance is primarily affected by the neutron lifetime and the nuclear rates of $^1\rm{H}(n, \gamma)\rm{D}$, $\rm{D}(d, n)^3\rm{He}$, and $\rm{D}(d, p)^3\rm{H}$~\cite{Pitrou2018PrecisionHe4predictions}. The latter two reactions also contribute to the deuterium uncertainty, together with \as{$\rm{D}(p, \gamma)^3\rm{He}$}~\cite{Pitrou2021ResolvingConclusions}, which has recently been improved by the LUNA collaboration~\cite{Mossa2020LUNArate}. 

\subsubsection{BBN numerical codes}
\label{sec:Numerical codes}
The most widely used numerical codes for cosmological analyses, which solve the aforementioned set of differential equations, are \parth\footnote{\url{http://parthenope.na.infn.it}}~\cite{Pisanti:2007hk, Consiglio:2017pot, Gariazzo:2021iiu} and \primat\footnote{\url{https://www2.iap.fr/users/pitrou/primat.htm}}~\cite{Pitrou2018PrecisionHe4predictions, Pitrou:2020etk}. Recently,  \prym\footnote{\url{https://github.com/vallima/PRyMordial}}~\cite{Burns:2023sgx} was released, which allows a simple implementation of NP.
Their theoretical predictions for the nuclear abundances differ slightly, mainly due to different implementations of the nuclear network rates. Other BBN codes, not considered here, are \texttt{AlterBBN}~\cite{Arbey:2018zfh}, and \texttt{LINX}~\cite{Giovanetti:2024zce}.

While different numerical code approaches for the weak rates represent a minor source of discrepancies, with differences at most of $0.2\%$~\cite{Pitrou2021ResolvingConclusions}, the choice of nuclear rates is undoubtedly crucial. The improved measurements of the deuterium burning rate~\cite{Mossa2020LUNArate} in \as{$\rm{D}(p, \gamma)^3\rm{He}$}, have refined the precision of the theoretical deuterium abundance and have led to a mild tension in the BBN results. The theoretical predictions of \primat\ as a function of the baryon-to-photon ratio ($\eta$) leads to a two standard deviations tension with the CMB determination~\cite{Pitrou:2020etk, Pitrou2021ResolvingConclusions}, questoning the widely assumed concordance between expectations and observations. Conversely, the predictions of \parth\ are still in complete agreement with the CMB data~\cite{Pisanti_deuteriumafterLUNA, Yeh_newdeuteriumrates}. This discrepancy is a consequence of the different adoptions for the nuclear rates and requires new nuclear data to settle the question. 

In light of this discrepancy, the numerical code \prym\ allows the user to choose between the PRIMAT nuclear rates, or the NACRE II nuclear rates reported in Ref.~\cite{NACREIIrates}, concordant with \parth\ predictions.
Additionally, the both aforementioned methods for normalising the weak rates can be selected, either through the experimental neutron lifetime or by involving the experimental values of $G_F$, $g_A$ and $V_{ud}$ (eq.~\ref{eq:K_theoretical}). Following the neutron lifetime prescription, the theoretical uncertainty for the PRIMAT (NACRE II) nuclear rates is given by $\sigma_{\Yp} = 1.1 \times 10^{-4}$ ($\sigma_{\Yp} = 1.4 \times 10^{-4}$) and $\sigma_{D/H} = 2.6 \times 10^{-7}$ ($\sigma_{D/H} = 1.0 \times 10^{-6}$). Therefore, the relative uncertainty for the helium-4 predictions is considerably lower than that for deuterium. Adopting the PRIMAT rates improves the precision for deuterium by a factor 4.

The resolution strategy of \prym\ is based on the philosophy of \primat, and consists of three stages. Firstly, the thermodynamic background is efficiently computed with \nudec, as described in Section~\ref{sec:Neutrino decoupling}, then the weak rates for the $n \leftrightarrow p$ conversion are calculated and lastly the nuclear abundances are solved. The flexibility of \prym, allowing to handle numerous details of the computation with boolean flags and to introduce NP quite easily, is the reason why we have chosen this tool in order to perform our analyses, see Section~\ref{sec:Results}. First, let us revisit the different BSM scenarios that will be implemented in the early universe in the following section.

\section{New Physics in the early universe}
\label{sec:NP}

\subsection{Non-standard neutrino interactions}
\label{sec:NSI}
Non-standard neutrino interactions (NSI) is a broad phenomenological framework that contains a plethora of NP models that may introduce additional particles with the ultimate objective of explaining the mechanism behind the mass generation of neutrinos~\cite{Dev2019, Farzan2017, Davidson:2003ha}. As an effective description of unknown physics operating at energies higher than the electroweak scale, the interaction vertex can be described as a contact interaction.  In general terms, NSI govern interactions between neutrinos and any fermion, but here we will restrict to two particularly relevant scenarios for the early universe: neutrino NC-NSI with electrons and neutrino CC-NSI with quarks. We will begin by focusing on the neutrino NC-NSI with electrons, governed by the following Lagrangian:
\begin{equation} \label{eq:NC NSI Lagrangian}
    \mathcal{L}_{\rm NSIe}^{\rm NC} = -2\sqrt{2}G_F \sum_{X, \alpha, \beta} \varepsilon_{\alpha\beta}^X ( \Bar{\nu}_\alpha \gamma^\mu P_L \nu_\beta ) (\Bar{e} \gamma_\mu P_X e ),
\end{equation}
following the usual notation where $\gamma^\mu$ are the Dirac matrices, $X = {R, L}$ represents the chirality so that $P_{R, L} = (1 \pm \gamma_5)/2$ are the chiral projectors and $\varepsilon_{\alpha\beta}^X$ are the strength of the NC-NSI relative to the electroweak one, connecting $\nu_\alpha$ and $\nu_\beta$, with $\alpha, \beta = e, \mu, \tau$ the flavour indices. If $\varepsilon^X_{\alpha\alpha} - \varepsilon^X_{\beta\beta} \neq 0$ the lepton flavour universality is violated. These type of interactions are called non-universal NSI. On the other hand, if $\varepsilon^X_{\alpha\beta} \neq 0$ when $\alpha \neq \beta$, the lepton flavour symmetry is no longer a conserved quantity. These interactions are known as flavour-changing NSI. 

In the context of the early universe, some combinations of neutrino NC-NSI with electrons are first introduced in Ref.~\cite{ManganoNSI}, and later updated in Ref.~\cite{deSalas:2016ztq}, although a more systematic approach is presented in Ref.~\cite{Stefano2021}. All of these works employ the density matrix evolution and find that the \Neff\ parameter varies in presence of NC-NSI due to the collision terms that describe neutrino-electron interactions and the contribution from neutrino oscillations in matter. \Neff\ is found to vary at the level of $\mathcal{O}(10^{-2})$, with the latter effect being completely negligible. This is a critical point, as the simplified neutrino decoupling presented in Section~\ref{sec:Neutrino decoupling} neglects neutrino oscillations. Within this approach, NC-NSI only alter the energy transfer rates, $\delta \rho_{\nu_\alpha}/\delta t$, increasing or decreasing the momentum-dependent distortions induced by the electron-positron annihilation. Accordingly, the SM couplings $g_{\alpha X}$ explicitly stated in eq.~\eqref{eq:couplings} are now a function of the NC-NSI parameters:  
\begin{equation}
    g_{\alpha X}^2 \longrightarrow \left( g_{\alpha X}^{\rm SM} + \varepsilon_{\alpha\alpha}^X \right)^2 + \sum_{\beta \neq \alpha} \abs{\varepsilon_{\alpha\beta}^X}^2.
\label{eq:BSM_couplings}
\end{equation}
From these expressions, the minimum value for \Neff~is found when $\varepsilon_{\alpha\alpha}^X = - g_{\alpha X}^{\rm SM}$ and $\varepsilon_{\alpha\beta}^X = 0$, since they minimise the energy transfer rates~\cite{Stefano2021}.

The variation in \Neff\ induced by neutrino NC-NSI with electrons is comparable to the near future observational precision. Therefore, cosmology will be able to constrain the strength of these interactions to be of the same order of magnitude as the weak interactions~\cite{Stefano2021}. Terrestrial experiments already constrain NC-NSI to be less strong than the weak interactions, $\varepsilon^X_{\alpha \beta} \lesssim \mathcal{O}(1-10^{-2})$, in a great variety of experimental setups. Neutrino oscillations~\cite{Bolanos:2008_NSI_solar_and_reactor, Agarwalla:2012_NSI_Borexino}, neutrino scattering~\cite{TEXONO:2010tnr}, as well as accelerator data~\cite{Bolanos:2008_NSI_solar_and_reactor, Barranco:2007ej} offer the most stringent current bounds, summarised in Ref.~\cite{Farzan2017} and compiled in the first two columns of Table~\ref{tab:NPbounds}.
The derivation of these constraints is typically conducted by taking one parameter at a time, or combining two of them at most.
Considering several free parameters at the same time activates degeneracies that have the immediate consequence of complicating the numerical calculations and worsening the limits significantly.
However, the bounds are more robust when considering multiple parameters at the same time. Studying cosmological constraints in addition to terrestrial ones allows to have complementary probes, for which the parameter degeneracies are different, thus reducing the impact of degeneracies on the final constraints.

\quad

Alternatively, neutrino CC-NSI with quarks affect the neutron-to-proton conversion by~\cite{Biggio:2009_NSIBounds}
\begin{equation}\label{eq:CCq NSI Lagrangian}
    \mathcal{L}^{\rm CC}_{\rm NSIq} = -2\sqrt{2}G_F V_{ud} \sum_{\alpha}\varepsilon^{udV}_{e\alpha} \left( \bar{u} \gamma_{\mu} P_L d \right) \left( \bar{e} \gamma^{\mu} P_L \nu_\alpha \right) + \rm h.c..
\end{equation}
where $V_{ud}$ is the relevant CKM matrix element, and the vector combination of CC-NSI parameters $\varepsilon^{udV}_{\alpha\beta}$ is defined as $\varepsilon^{udV}_{\alpha\beta} \equiv \varepsilon^{udR}_{\alpha\beta} + \varepsilon^{udL}_{\alpha\beta}$.
The modifications to the neutron beta-minus decay, and consequently, to all interactions maintaining neutrons and protons in chemical equilibrium until the weak interactions freeze-out, are given by~\cite{Santos:2020dgs}
\begin{equation}
\label{eq:CC-NSI factor}
    \Gamma^{\rm obs}_\beta = \Gamma_\beta \left( 1 + 2 \Re(\varepsilon^{udV}_{ee}) + \sum_\alpha \abs{\varepsilon^{udV}_{e\alpha}}^2 \right),
\end{equation}
where $\Gamma_\beta$ is the predicted rate in the SM. Provided that neutrino flavours have different temperatures, the neutron beta decay rate consists of two separate contributions when assuming that $T_{\numu} = T_{\nutau}$. 
Accordingly, the weak rates in eq.~\eqref{eq:np conversion} are now given by 
\begin{equation}
\label{eq:CC-NSI weak rates}
    \Tilde{\Gamma}_{n \rightarrow p} = \Gamma_{n \rightarrow p}(T_{\nue}) \left( 1 + 2\Re{\varepsilon^{udV}_{ee}} + \abs{\varepsilon^{udV}_{ee}}^2 \right) + \Gamma_{n \rightarrow p}(T_{\nu_{\mu, \tau}}) \left( \abs{\varepsilon^{udV}_{e\mu}}^2 +\abs{\varepsilon^{udV}_{e\tau}}^2 \right) ,
\end{equation}
where the first term is associated with CC interactions such as those present in the SM, and the second term replaces \nue\ in the weak interactions in eq.~\eqref{eq:weak interactions} with \numu\ or \nutau. An analogous expression holds for the $p \rightarrow n$ rates.
The strength of neutrino CC-NSI with quarks has been found to be of the order of $\mathcal{O}(10^{-2}- 10^{-4})$~\cite{Biggio:2009_NSIBounds} compared to that of the weak interactions, thus being even more constrained than neutrino NC-NSI with electrons. For this reason, they have not yet been studied in the context of BBN~\cite{ManganoNSI}. The improvement in the precision of the observed abundances has prompted this work to now investigate the potential impact of neutrino CC-NSI with quarks on BBN.

\subsection{Non-unitary three-neutrino mixing}

Another potential explanation for neutrino masses, and a consequence of a plethora of BSM theories, invokes the existence of hypothetical heavy neutral leptons (HNLs). The active three-flavour neutrinos ($\nu_\alpha$) present in the SM would be mixed with $n$ possible mass eigenstates ($\nu_i$), including those corresponding to the HNLs. The $n \times n$ mixing matrix, $K$, would lead to modifications of the neutrino NC and CC interactions, described in the low energy limit by~\cite{Escrihuela:2015wra, Escrihuela:2016ube}:
\begin{equation}
    \mathcal{L}_{\rm CC} = -2\sqrt{2}G_F \sum_{i, j}  ( K^\dag) _{ie} K_{ej} (\bar{\nu}_i \gamma^\mu P_L \nu_j) (\bar{e} \gamma_\mu P_L e),
\end{equation} 
\begin{equation}
    \mathcal{L}_{\rm NC} = -2\sqrt{2}G_F \sum_{X = L, R} g_X \sum_{i, j}  ( K^\dag K)_{ij} (\Bar{\nu}_i \gamma^\mu P_L \nu_j) (\Bar{e} \gamma_\mu P_L e),
\end{equation}
where indices $i$ and $j$ represent the mass eigenstates and go from $1$ to $n$. The sum over mass eigenstates is limited to the heaviest kinematically accessible eigenstate.

The full $n \times n$ lepton mixing matrix can be decomposed into two blocks, $K=(N \ S)$. The first one, $N$, relates the three lightest states, while the second, $S$, describes the mixing between the three lightest states and the remaining $n-3$ heavier states. This decomposition is particularly useful in the low-energy limit (the SM energy scale), as only $N$ is relevant in this regime. Even if $K$ is generally assumed to be unitary, the blocks themselves are not required to be unitary. The deviations of $N$ from unitarity are parametrised through the coefficients $\alpha_{ij}$,
\begin{equation}
     N = \begin{pmatrix}
    \alpha_{11} & 0 & 0 \\
    \alpha_{21} & \alpha_{22} & 0 \\
    \alpha_{31} & \alpha_{32} & \alpha_{33} 
    \end{pmatrix} U_{\rm PMNS}, 
\end{equation}
where $U_{\rm PMNS}$ is the standard unitary leptonic mixing matrix. Note that in the SM, $\alpha_{ij} = \delta_{ij}$, that is to say, diagonal entries are 1 and off-diagonal elements are null. The diagonal parameters, $\alpha_{ii}$, are real, whereas the non-diagonal ones, $\alpha_{ij}$ ($i \neq j$), could be complex, contributing to CP violation. Diagonal and non-diagonal parameters are related through triangular inequalities:
\begin{equation}
\label{eq:unitarity condition}
    \alpha_{ij} \leq \sqrt{(1 - \alpha_{ii}^2) (1 - \alpha_{jj}^2)}.
\end{equation}
If deviations from unitarity are small enough we can establish a correspondence between the NU parameters and the previously presented NC-NSI ones~\cite{Gariazzo2022Non-unitarity}. Otherwise, it is mandatory to do the full calculation in order to take into account only kinematically accessible states. The following relations connect them:
\begin{equation}
\begin{aligned}
    \varepsilon^{L}_{\alpha\beta} = -(\delta_{\beta e} \delta_{\alpha e} + g_L \delta_{\alpha\beta}) &+ \frac{(NN^\dagger)_{\alpha e} (NN^\dagger)_{e\beta} + g_L (NN^\dagger)^2_{\alpha\beta}}{\sqrt{\alpha_{11}^2 (\alpha_{22}^2 + |\alpha_{21}|^2)}} ,\\ 
    \\
    \varepsilon^{R}_{\alpha\beta} = -g_R \delta_{\alpha\beta} &+ \frac{g_R (NN^\dagger)^2_{\alpha\beta}}{\sqrt{\alpha_{11}^2 (\alpha_{22}^2 + |\alpha_{21}|^2)}} .
    \label{eq:NUtoNSI}
\end{aligned}
\end{equation}
These relations also include the influence of non-unitarity (NU) on the Fermi constant, which is modified in presence of HNLs. Particularly, the Fermi constant measured in the muon decay, $G_F^\mu$, is related to the actual Fermi constant, $G_F$, by~\cite{Escrihuela:2015wra}
\begin{equation}
    G_F^\mu = G_F \sqrt{(N^\dag N)_{ee} (N^\dag N)_{\mu\mu}} = G_F \sqrt{\alpha_{11}^2(\alpha_{22}^2 + \abs{\alpha_{21}}^2)} ,
\label{eq:G_Fmu}
\end{equation}
while the beta decay is also altered
\begin{equation}
    G_F^\beta = G_F \sqrt{(N N^\dag)_{ee} } = G_F \alpha_{11} = \frac{G_F^\mu}{\sqrt{\alpha_{22}^2 + \abs{\alpha_{21}}^2}} ,
\label{eq:G_Fbeta}
\end{equation}
and we will adopt the muon decay Fermi constant since it is the most precise measurement, $G_F^\mu = 1.1663787(6) \times 10^{-5}~\rm{GeV}^{-2}$~\cite{ParticleDataGroup2022}. 
Thus, the impact of NU on the early universe is twofold. First, it affects the thermodynamic background due to the alterations to NC and CC interactions, synthesised in its mapping to neutrino NC-NSI with electrons. But second, the change in the beta decay affects the neutron-to-proton conversion and therefore the nuclear abundances. 

\begin{table}[h]
    \centering
    \hspace*{-0.04\textwidth}
    \setlength{\tabcolsep}{9pt}
    \begin{tabular}{@{}cccccc@{}}
        \toprule
        \multicolumn{4}{c}{\textbf{Neutrino NC-NSI with electrons (90\% CL)}} &  \multicolumn{2}{c}{\textbf{Neutrino CC-NSI with}} \\ \cmidrule(r){1-4} 
        \multicolumn{2}{c}{\textbf{Non-universal NSI}} &  \multicolumn{2}{c}{\textbf{Flavour-changing NSI}} & \multicolumn{2}{c}{\textbf{quarks (90\% CL)~\cite{Biggio:2009_NSIBounds}}}\\  
        \toprule
   
        \multicolumn{2}{c}{$-0.021 < \varepsilon^L_{ee} < 0.052$~\cite{Bolanos:2008_NSI_solar_and_reactor}} & \multicolumn{2}{c}{$-0.13 < \varepsilon^{L, R}_{e\mu} < 0.13$~\cite{Barranco:2007ej}} & \multicolumn{2}{c}{$\abs{\Re{\varepsilon^{udV}_{ee}}} < 8.6 \times 10^{-4}$} \\
        \multicolumn{2}{c}{$-0.07 < \varepsilon^R_{ee} < 0.08$~\cite{TEXONO:2010tnr}} & \multicolumn{2}{c}{$-0.33 < \varepsilon^L_{e\tau} < 0.33$~\cite{Barranco:2007ej}} & \multicolumn{2}{c}{$\varepsilon^{udV}_{e\alpha} < 0.041$} \\
        \multicolumn{2}{c}{$-0.03 < \varepsilon^{L, R}_{\mu\mu} < 0.03$~\cite{Barranco:2007ej}} & \multicolumn{2}{c}{$-0.28 < \varepsilon^R_{e\tau} < -0.05$~\cite{Barranco:2007ej}}
         & &\\
        \multicolumn{2}{c}{$-0.12 < \varepsilon^L_{\tau\tau} < 0.06$~\cite{Bolanos:2008_NSI_solar_and_reactor}} & \multicolumn{2}{c}{$0.05 < \varepsilon^R_{e\tau} < 0.28$~\cite{Barranco:2007ej}} & & \\
        \multicolumn{2}{c}{$-0.98 < \varepsilon^R_{\tau\tau} < 0.23$~\cite{Bolanos:2008_NSI_solar_and_reactor, Agarwalla:2012_NSI_Borexino}} & \multicolumn{2}{c}{$-0.19 < \varepsilon^R_{e \tau} < 0.19$~\cite{TEXONO:2010tnr}}
         & & \\
        \multicolumn{2}{c}{$-0.25 < \varepsilon^R_{\tau\tau} < 0.43$~\cite{Bolanos:2008_NSI_solar_and_reactor}} & \multicolumn{2}{c}{$-0.10 < \varepsilon^{L, R}_{\mu\tau} < 0.10$~\cite{Barranco:2007ej}} & & \\
        \bottomrule
        \toprule
    
        \multicolumn{6}{c}{\textbf{Non-unitary three-neutrino mixing (3$\sigma$ Bounds)~\cite{ForeroNUbounds}}} \\
        \toprule
       $\alpha_{11} > 0.93$ & $\alpha_{22} > 0.98$ & $\alpha_{33} > 0.72$  &  $\abs{\alpha_{21}} < 0.025$ & $\abs{\alpha_{31}} < 0.075$ & $\abs{\alpha_{32}} < 0.02$\\
        \bottomrule
    \end{tabular}
    \caption{Current bounds on neutrino NC-NSI with electrons, neutrino CC-NSI with quarks, and non-unitary three-neutrino mixing given by terrestrial experiments. Adapted from Refs.~\cite{Stefano2021, Gariazzo2022Non-unitarity, Biggio:2009_NSIBounds}.}
    \label{tab:NPbounds}
\end{table}

Neutrino decoupling is already examined in the context of NU in Ref.~\cite{Gariazzo2022Non-unitarity}, with a modified version of the \fortepiano~\cite{Gariazzo:2019gyi, Bennett:2020zkv} numerical code, adapted to operate in the mass basis and to account for the kinematically accessible eigenstates.
It was found that the main effect of non-unitarity on neutrino decoupling arises from the change in the Fermi constant entering the neutrino collision term. 
Significant departures from unitarity parameterised by $\alpha_{11}$ and $\alpha_{22}$ lead to a delayed decoupling and increase \Neff. Since $G_F$ is independent of $\alpha_{33}$, \Neff~is rather insensitive to it, and $\alpha_{33}$ is left essentially unconstrained. Cosmological bounds for NU parameters were not found to be competitive with terrestrial bounds, mainly derived from oscillation experiments~\cite{ForeroNUbounds}, and reported in the last row of Table~\ref{tab:NPbounds}.

\quad

In summary, neutrino NC-NSI with electrons are expected to predominantly affect neutrino decoupling and thus the \Neff\ value, whereas neutrino CC-NSI are expected to solely affect the $n \leftrightarrow p$ conversion and thus the nuclear abundances. In turn, NU is expected to modify both cosmological observables.

\section{Results}
\label{sec:Results}

Firstly, we have developed a modified version of the \nudec~code to deal with the changes in the thermodynamic background prompted by neutrino NC-NSI with electrons and non-unitary three-neutrino mixing. We have checked that \nudec~approximations are valid even in such BSM scenarios, by comparing the results with the complete neutrino decoupling calculations presented in Refs.~\cite{Stefano2021, Gariazzo2022Non-unitarity}, made with \fortepiano. Thus, we can be certain that the \Neff~calculation is accurate, with uncertainties one order of magnitude smaller than the forecasted experimental precision from next-generation observations. Furthermore, we allow the neutrino fluid to be described by several temperatures, $T_{\nu_\alpha}$. After exploring different settings, we check that the scenario that finds a better agreement with the complete calculation is the one corresponding to two different temperatures, $T_{\nu_e} \neq T_{\nu_{\mu, \tau}}$, which better mimics the small effect of neutrino oscillations.

We adapted the BBN code \prym, which comprises the implementation of neutrino decoupling \textit{\'a la} \nudec, to accommodate the changes described above regarding the thermodynamic background but also to reflect possible changes in the normalisation of the weak rates. Such effects would be driven by: (i) neutrino CC-NSI with quarks, and (ii) non-unitarity, that shifts the Fermi constant through eq.~\eqref{eq:G_Fbeta}. 
Separately, we allow weak rates to account also for the two independent neutrino temperatures, $T_{\nue}$ and $T_{\nu_{\mu, \tau}}$, following eq.~\eqref{eq:CC-NSI weak rates}. 

Lastly, we use the observational measurements of the light element abundances presented in Ref.~\cite{ParticleDataGroup2022}: $\rm{D/H} = (2.547 \pm 0.025) \times 10^{-5} , Y_P = 0.245 \pm 0.003, ^7\rm{Li/H} = (1.6 \pm 0.3) \times 10^{-10}$. The EMPRESS determination of helium-4~\cite{Matsumoto:2022tlr}, which is $\sim 1 \sigma$ lower than the rest, and points to a potential lepton asymmetry~\cite{Burns:2022hkq}, is not considered here.
We will also use the Milky Way estimate of $^3\rm{He}/\rm{H} = (0.9 - 1.3) \times  10^{-5}$~\cite{Bania:2002yj}, recalling that the helium-3 determination lacks cosmological significance because it is unclear whether it truly reflects a primordial abundance~\cite{Pitrou2018PrecisionHe4predictions}. It provides only an approximate indication of the expected order of magnitude. Similarly, lithium-7 measurements are in stark conflict with the SBBN prediction and this is not typically resolved consistently in BSM scenarios. Nevertheless, for the sake of completeness, the results on both abundances will be shown, except where statistical treatments are considered.

Results are obtained with the large nuclear network available in \prym~for more precise results. The remaining parameters such as the baryon-to-photon ratio are chosen to be consistent with the CMB determination. In case the baryon-to-photon ratio is left free, we expect all limits to be relaxed with respect to our current findings, because of the additional degree of freedom. It is however difficult to justify ignoring the CMB limits on such quantity, since they provide the strongest constraints on the amount of baryons in the universe available to date.

\subsection{BBN and NSI}

\begin{figure}
    \centering
    \includegraphics[width = 0.9\textwidth]{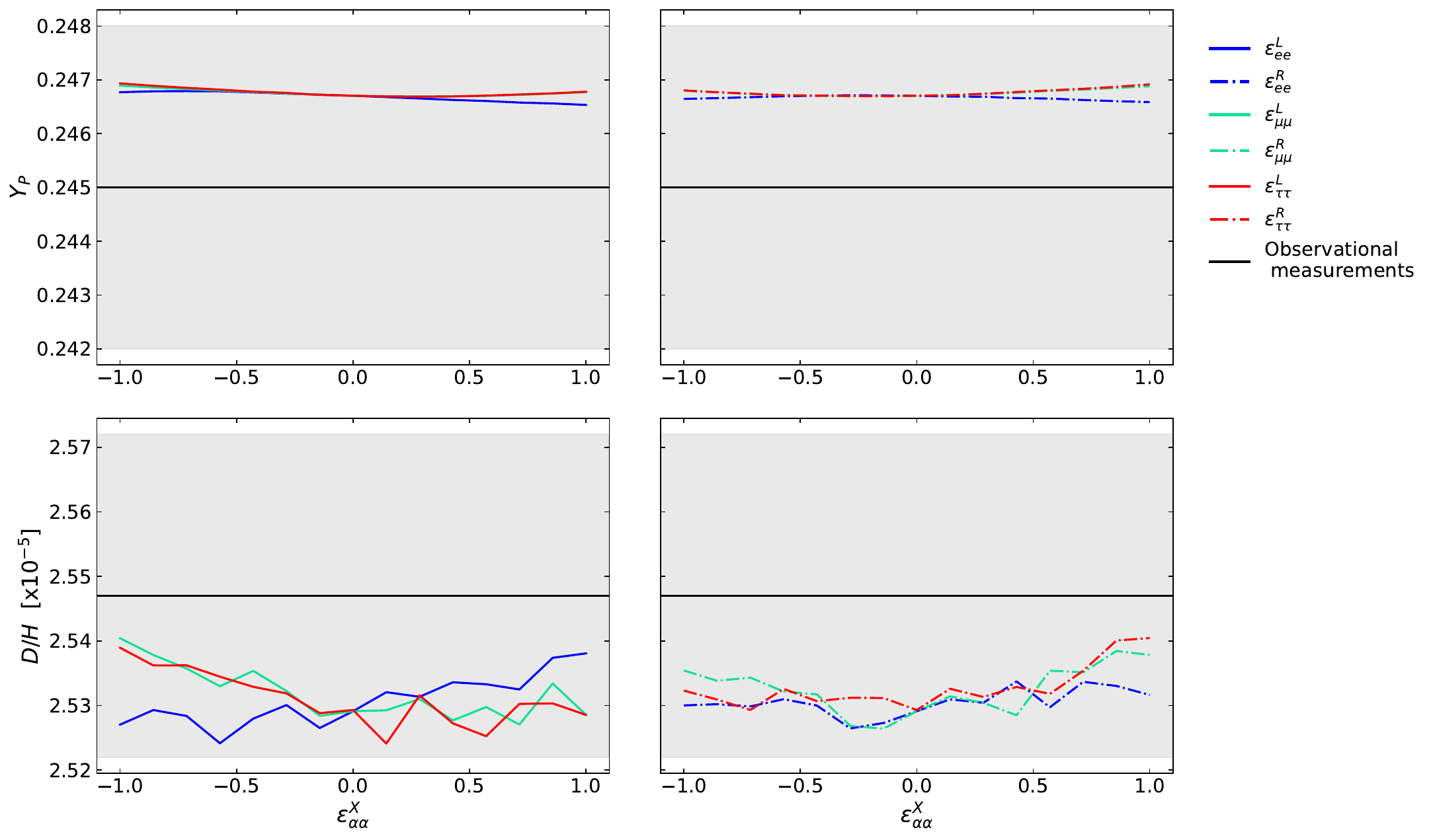}
    \caption{Helium-4 (top panel) and deuterium (bottom panel) abundances as a function of neutrino NC-NSI with electrons parameters. BBN is mainly insensitive to $\mathcal{O}(1)$ $\varepsilon^X_{\alpha \beta}$.}
    \label{fig:BBN_vs_NC-NSI}
\end{figure}

Neutrino NC-NSI with electrons result mainly in a different thermodynamic history, which is summarised by the changes in \Neff. Neutrino decoupling is significantly more sensitive to NC-NSI parameters than nuclear abundances, as already noted in Ref.~\cite{ManganoNSI} for certain combinations of $\varepsilon^X_{ee}$, $\varepsilon^X_{\tau \tau}$ and $\varepsilon^X_{e\tau}$. We check that the current precision of the observed nuclear abundances is insufficient to reject any NC-NSI configuration (see Figure~\ref{fig:BBN_vs_NC-NSI}). The corrections to the weak rates, although modified by the different neutrino and photon temperatures, also remain virtually unchanged.

\begin{figure}
    \centering
    \includegraphics[width = 0.9\textwidth]{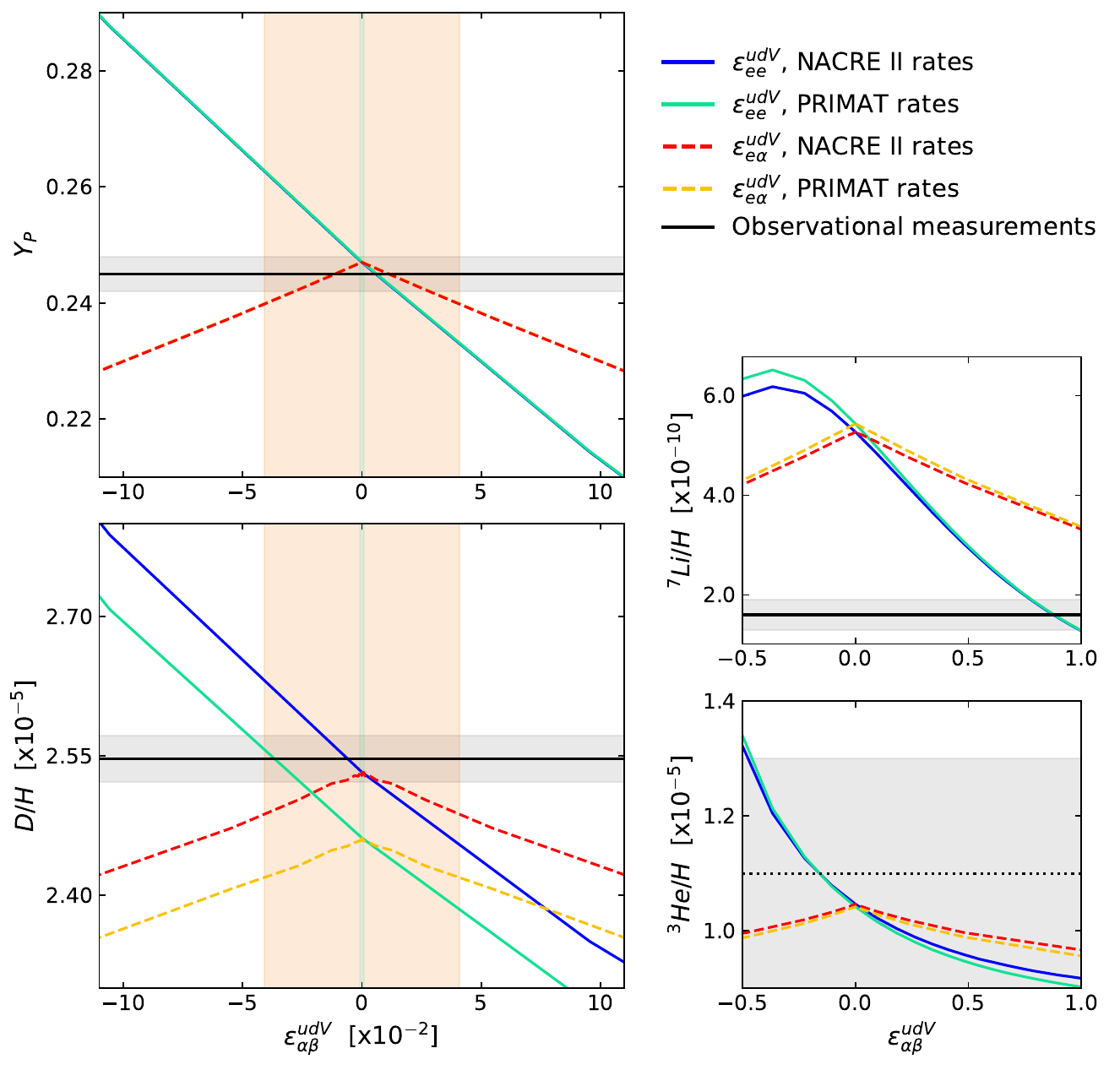}
    \caption{Nuclear abundances as a function of neutrino CC-NSI with quarks, mainly modifying neutron-to-proton conversion. Solid lines correspond to $\varepsilon^{udV}_{ee}$ and dashed lines to $\varepsilon^{udV}_{e\alpha}$. Blue and red lines are obtained with NACRE II rates, while green and yellow lines with PRIMAT rates.}
    \label{fig:BBN_vs_CC-NSI}
\end{figure}

In the case of neutrino CC-NSI with quarks, the situation is the opposite: all nuclear abundances are highly sensitive to the normalisation of the weak rates, in contrast to the thermodynamic background dependence. As previously stated in Section~\ref{sec:Numerical codes}, we must use the normalisation of the weak rates in eq.~\eqref{eq:K_theoretical}, because all the BSM physics affecting the neutron decay are reabsorbed if the experimental neutron lifetime is used instead~\cite{ManganoNSI}. The neutron decay width in eq.~\eqref{eq:CC-NSI weak rates} also incorporates both neutrino temperatures, with $T_{\nu_{\mu, \tau}}$ differing from $T_{\nu_e}$ by only $\sim 0.01-0.1\%$, so that the corresponding correction is insignificant. 

In Figure~\ref{fig:BBN_vs_CC-NSI}, the solid lines represent the nuclear abundances as a function of $\varepsilon^{udV}_{ee}$, while dashed lines represent the variation with $\varepsilon^{udV}_{e\alpha}$, and $\alpha \neq e$. NACRE II nuclear rates (in blue and red) are represented together with PRIMAT nuclear rates (in green and yellow), and show the previously mentioned tension for deuterium (lower left panel). Horizontal grey bands represent the current observed nuclear abundances and vertical coloured bands represent the bounds on CC-NSI parameters derived by terrestrial experiments. 

The structure of the factor multiplying the neutron beta decay width, presented in eq.~\eqref{eq:CC-NSI factor}, is responsible for the observed faster variation for $\varepsilon^{udV}_{ee}$ than for $\varepsilon^{udV}_{e\alpha}$. The term proportional to the real part of $\varepsilon^{udV}_{ee}$ permits the enhancement or suppression of the neutron beta decay, whereas the quadratic dependence on $\varepsilon^{udV}_{e\alpha}$ favours exclusively a faster neutron decay. This implies that nuclear abundances may exceed or fall below the SBBN prediction for $\varepsilon^{udV}_{ee}$ while being necessarily lower for $\varepsilon^{udV}_{e\alpha}$. In the latter case, the maximum value, which coincides with the SBBN prediction is reached when $\varepsilon^{udV}_{e\alpha} = 0$, i.e., in the absence of CC-NSI that relate different flavours. If the observations were accurate enough and indicated values exceeding the SBBN prediction, it would be possible to rule out the presence of this type of CC-NSI alone. 

Note the different scale, helium-4 and deuterium abundances (left panels) are able to constrain CC-NSI parameters to the order of $\mathcal{O}(10^{-2})$. In fact, these constraints are competitive with terrestrial bounds, especially for $\varepsilon^{udV}_{e\alpha}$ (orange vertical band), though for $\varepsilon^{udV}_{ee}$ the terrestrial limits are more stringent, of $\mathcal{O}(10^{-4})$ (turquoise vertical band). Helium-3 and lithium vary at a much slower rate (right panels) and are accordingly represented on a different scale. The lithium problem is solved for ruled out CC-NSI parameters, at the cost of hugely underproducing deuterium and helium-4.

\subsection{BBN and non-unitary three-neutrino mixing}

\begin{figure}
    \centering
    \includegraphics[width = 0.65\textwidth]{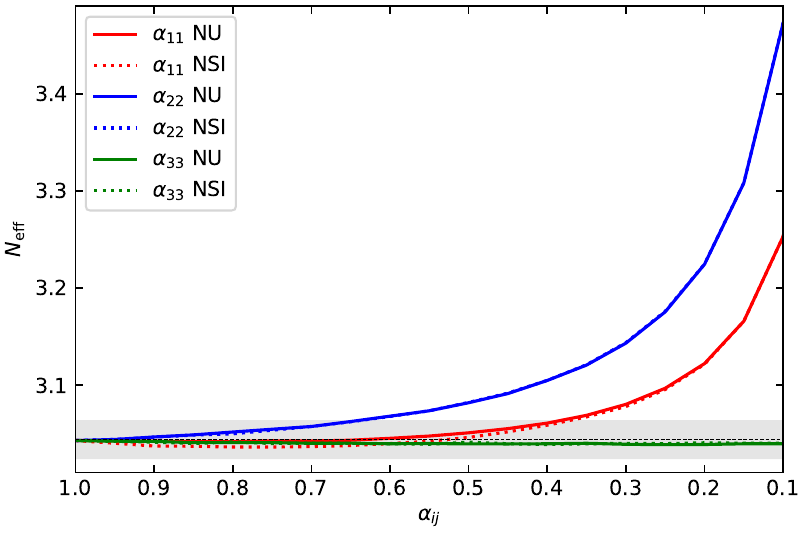}
    \caption{\Neff~as a function of NU parameters, when the exact calculation is made with \fortepiano~(solid lines) and when we use the approximate mapping in eq.~\eqref{eq:NUtoNSI} (dashed lines). Note the slight discrepancies ($\sigma (\Neff) \lesssim 0.006$) for $\alpha_{11} > 0.5$.}
    \label{fig:NU_desv}
\end{figure}

In principle, the correspondence between NU and neutrino NC-NSI with electrons in eq.~\eqref{eq:NUtoNSI} allows for the straightforward computation of NU effects in neutrino decoupling, avoiding the exact description in terms of the neutrino mass eigenstates, $\nu_i$.  The \nudec\ approach finds agreement with the full computation~\cite{Gariazzo2022Non-unitarity} for $\alpha_{22}$ and $\alpha_{33}$. Conversely, for $\alpha_{11}$, there are minor differences of the order of $0.2 \%$ on \Neff~(see Figure~\ref{fig:NU_desv}), which appear also when \fortepiano~is ran using the approximate mapping between NU and NC-NSI. Then, the inconsistencies found in \nudec\ are not attributable to the neutrino decoupling approximations, but rather to the inexact relation that links NU and NC-NSI. Still, such discrepancies are not yet large enough to be relevant when compared to the expected observational precision for \Neff.

\begin{figure}
    \centering
    \includegraphics[width = 0.9\textwidth]{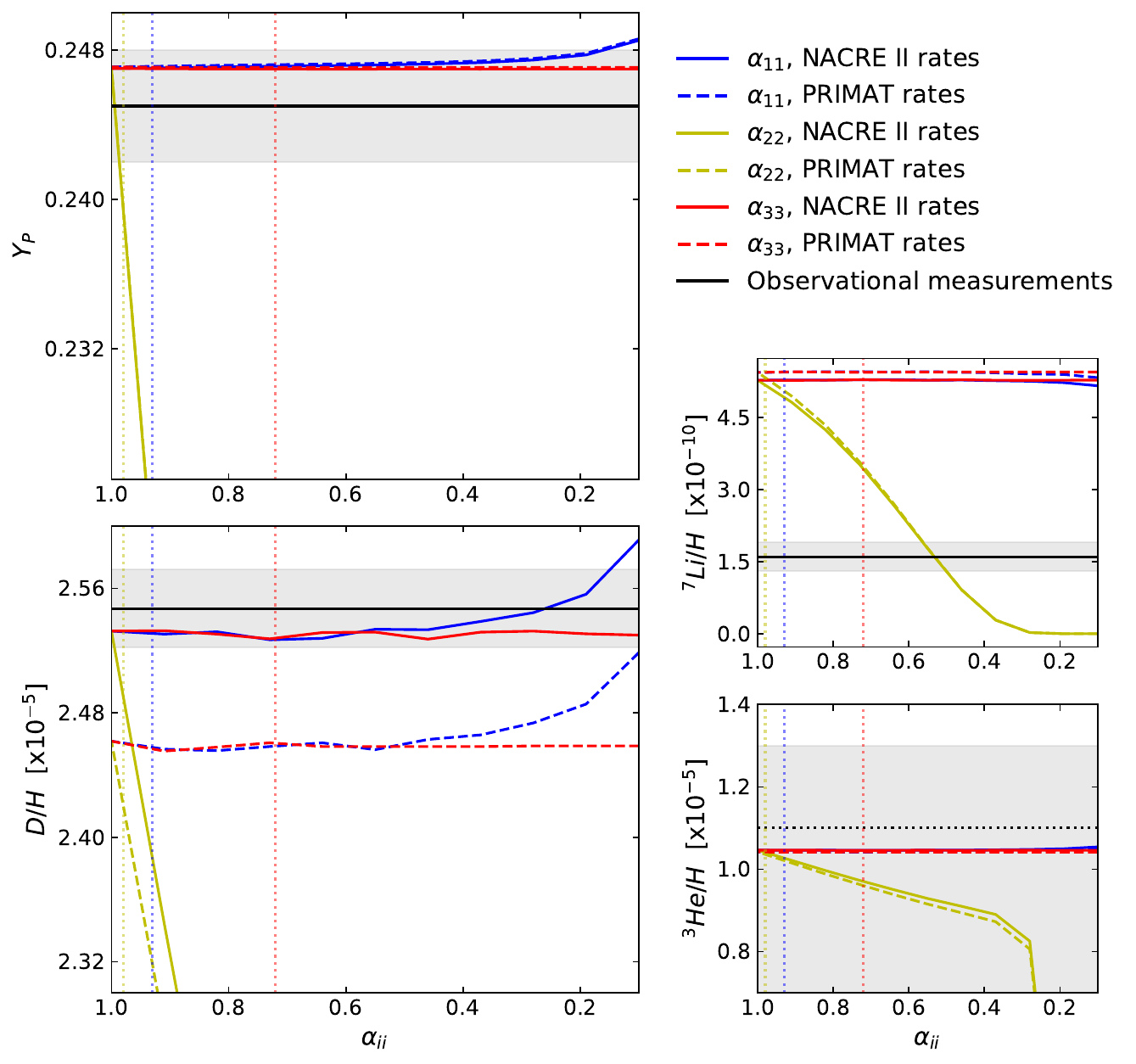}
    \caption{Nuclear abundances as a function of NU diagonal parameters: $\alpha_{11}$ (blue lines), $\alpha_{22}$ (yellow lines) and $\alpha_{33}$ (red lines). Results are obtained with NACRE II rates (solid lines) and PRIMAT rates (dashed lines).}
    \label{fig:BBN_vs_alpha_ii}
\end{figure}

As previously noted, alterations to neutrino decoupling are not as determinant for BBN abundances as the normalisation of the weak rates, which instead plays a crucial role. Accordingly, we expect non-unitarity to affect BBN mainly through the modified beta decay Fermi constant in eq.~\eqref{eq:G_Fbeta}, rather than through its impact on neutrino decoupling. $G_F^\beta$ depends only on $\alpha_{22}$ and $\alpha_{21}$, but it does not depend on $\alpha_{11}$, as opposed to the shifted $G_F$ in eq.~\eqref{eq:G_Fmu}, which enters in the collision term. In Figure~\ref{fig:BBN_vs_alpha_ii}, the BBN yields associated to both nuclear rates are shown as a function of the NU diagonal parameters. Non-unitarity gives rise to two fundamental effects that operate concurrently: the change in \Neff\ already studied in Ref.~\cite{Gariazzo2022Non-unitarity}, and the mismatch between $G_F^\beta$ and $G_F^\mu$. The nuclear abundances are strongly dependent on $\alpha_{22}$ (yellow lines) because of its effect on $G_F^\beta$. The fact that $G_F^\beta \geq G_F^\mu$ explains why the abundances are diminished, because the neutron-to-proton conversion is enhanced. Conversely, $\alpha_{11}$ (blue lines) presents a mild dependence, as a result of the \Neff\ increase, and $\alpha_{33}$ (red lines) has practically no influence on the results.

Just like for the neutrino decoupling, BBN yields are mostly sensitive to $\alpha_{22}$. Strikingly, BBN imposes particularly stringent constraints on $\alpha_{22}$; not only does it enhance those derived from neutrino decoupling, but it also slightly outperforms those obtained from terrestrial experiments (vertical dotted lines).

To further study the effect of the joint action of multiple NU parameters, we follow the approach of Refs.~\cite{Gariazzo2022Non-unitarity, Consiglio:2017pot} and introduce the following $\chi^2$ function
\begin{equation}
    \chi^2 (\Vec{\alpha}_{ij}) = \frac{[X^{\rm th}(\Vec{\alpha}_{ij}) - X^{\rm exp} ]^2}{\sigma^2_{\rm exp}(X) + \sigma^2_{\rm th}(X)}  ,
\end{equation}
where $X$ is a cosmological observable, either \Neff\ or the nuclear abundances of deuterium or helium-4. The set of adopted NU parameters is represented by $\Vec{\alpha}_{ij}$. It is assumed that $(\Neff)^{\rm exp} = 3.044$ matches the SM value and that $\sigma_{\rm exp}(\Neff) = 0.02$ is the predicted future uncertainty. The numerical uncertainty on \Neff\ is not included, as it is subdominant. For the nuclear abundances, $X^{\rm exp}$ is taken to be the observations reported in Ref.~\cite{ParticleDataGroup2022} and the theoretical uncertainties for \prym, $\sigma_{\rm th}(X)$, are given in Section~\ref{sec:Numerical codes} and depend on the nuclear rates~\cite{Burns:2023sgx}. Note that they are actually evaluated for the normalisation of the weak rates with $\tau_n^{\rm exp}$. For the sake of rigour, it would be necessary to perform an analysis of the theoretical uncertainties to guarantee that they remain unchanged even with our different prescription for the normalisation. Nonetheless, we expect the uncertainties to increase slightly, yet not substantially.

\begin{figure}
    \centering 
    \hspace*{-0.1\textwidth}
    \includegraphics[width=1.2\textwidth]{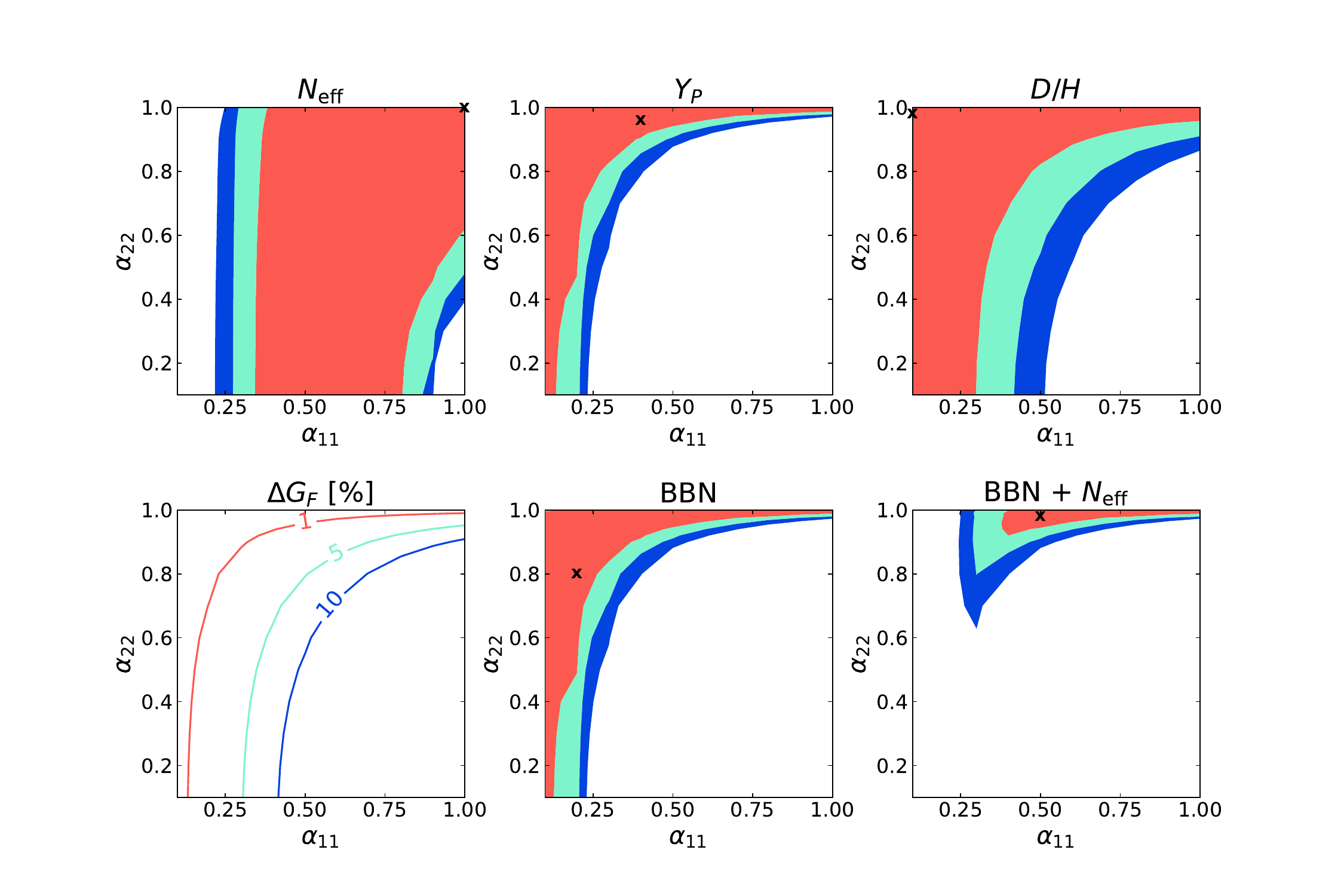}
    \caption{Allowed $1\sigma$ (red), $2\sigma$ (turquoise) and $3\sigma$ (blue) regions on the $\alpha_{22}$-$\alpha_{11}$ plane. Upper panels: \Neff, helium-4 and deuterium abundances constraints. Lower panels: $\Delta G_F$ minimum values and BBN constraints alone ($Y_p$ and $D/H$) and in combination with \Neff. It has been profiled over $\alpha_{21}$ values, satisfying the unitarity condition, eq.~\eqref{eq:unitarity condition}. 
    In the case of \Neff\ we consider the future measurements, while for BBN we consider the current ones.
    The crosses mark the best-fit for each case; further details can be found in the main text.}
    \label{fig:alpha_plane_NACRE}
\end{figure}

Given that the principal source of sensitivity stems from changes in the Fermi constant, $\alpha_{11}$, $\alpha_{22}$ and $\alpha_{21}$ are the most relevant NU parameters. 
Hence, we compute \Neff\ and the nuclear abundances with the NACRE II rates, in a numerical grid for $\alpha_{11}$, $\alpha_{22}$ and $\alpha_{21}$, storing the results and subsequently calculating the $\chi^2$. Then, the global minimum is found, $\chi_{\rm min}^2$, and the difference $\Delta\chi^2 = \chi^2 - \chi_{\rm min}^2$ is the quantity of interest. Figure~\ref{fig:alpha_plane_NACRE} shows allowed regions for $\alpha_{11}$ and $\alpha_{22}$ at $1$, $2$ and $3\sigma$, after profiling over $\alpha_{21}$, which is subject to follow the unitarity condition presented in eq.~\eqref{eq:unitarity condition}. The upper panels display the helium-4 (center) and deuterium (right) abundances, while the \Neff\ panel (left) corresponds to Ref.~\cite{Gariazzo2022Non-unitarity} and is shown for comparison. The lower left panel shows the percentage change in the Fermi constant for the beta decay, $G^\beta_F$, with respect to that measured in the muon decay $G^\mu_F$. The quantity $\Delta G_F = (G^\beta_F - G^\mu_F)/G^\mu_F$ parametrises how NU affects the neutron-to-proton conversion. The remaining two lower panels are the combinations of BBN abundances (center), and its combination with \Neff\ (right). The crosses mark the best-fit point for each case. The SM is recovered when $\alpha_{ij} = \delta_{ij}$, that is, corresponds to the upper right corner for all panels.

Firstly, let us focus on nuclear abundances. The helium-4 regions are more restrictive than those of deuterium, but this is due to the lower precision in deuterium theoretical computations with the NACRE II rates. Both of them exhibit a pretty similar behavior, and the same trend is also evidenced in the lower left panel, which plots $\Delta G_F$. This reinforces the conclusion that the BBN yields are mainly sensitive to changes in the normalization of the weak reactions rather than to any other effect. In contrast with the shape of the regions, we previously observed that the nuclear abundances should be nearly independent of $\alpha_{11}$, given that $G^\beta_F$ is also independent. This apparent contradiction can be explained by the indirect dependence that arises from the fact that $\alpha_{21}$, which does influence $G^\beta_F$, is limited by $\alpha_{11}$ through the inequality imposed from the unitarity of the full neutrino mixing matrix (eq.~\eqref{eq:unitarity condition}). In fact, $\Delta G_F$ is found to be minimal when $\alpha_{21}$ reaches its maximum value, $\abs{\alpha^{\rm max}_{21}} = \sqrt{( 1 - \alpha_{11})^2 ( 1 - \alpha_{22})^2}$, partially compensating the deviations from unitarity of $\alpha_{22}$. For the nuclear abundances, in general, the minimum value of the $\chi^2$ function is found for $\alpha^{\rm max}_{21}$, except for a small region in the upper left corner. For these parameters, \Neff\ is relatively large ($\sim 3.2$), appreciably increasing the nuclear abundances. This can only be balanced by a larger value of $G^\beta_F$ that favors a more efficient neutron decay. In summary, the profiling of $\alpha_{21}$ enables the establishment of a trade-off between \Neff\ and the beta decay Fermi constant. While in general the non-unitarity leads to the minimization of $\Delta G_F$ as the dominant effect in BBN, for configurations where \Neff\ is particularly large (left side of the $\alpha_{22}$-$\alpha_{11}$ plane) both aspects become relevant. 
The remaining two panels show the combination of both helium-4 and deuterium abundances, as well as these BBN abundances together with \Neff. Such combinations are obtained by first calculating the $\chi^2 (\alpha_{11}, \alpha_{22}, \alpha_{21})$ separately for each observable and then summing the values for the desired combination. Then, the minimum value is found for each pair $\{ \alpha_{11}, \alpha_{22} \}$. In all combinations, the $\alpha_{21}$ values typically correspond to $\alpha^{\rm max}_{21}$, meaning that, in general terms, minimising $\Delta G_F$ is the most paramount aspect once again. 
We find that the more precise helium-4 theoretical calculations constitute the dominant contribution to the BBN results, which manifest a notable improvement in their constraining power with respect to \Neff\ results (upper left panel). This improvement is even more remarkable considering that \Neff\ uncertainties represent the future precision, whereas the BBN observations are the current measurements. In any case, the two regions are complementary, and when the BBN abundances are combined with \Neff, the parameter space is constrained to a much greater extent than would be possible with either observable alone. This is an excellent demonstration of the synergies that can be achieved between neutrino decoupling and light nuclei production. Both tests can be used together to break degeneracies and provide stronger constraints on BSM physics scenarios. It is remarkable that the combination of all observables indicates a slight preference for deviations from unitarity ($\alpha_{ii} = 1$), as evidenced by the best-fit point. Despite this, the SM always remains within the 1$\sigma$ region.

\begin{figure}
    \centering
    \includegraphics[width = 0.75\textwidth]{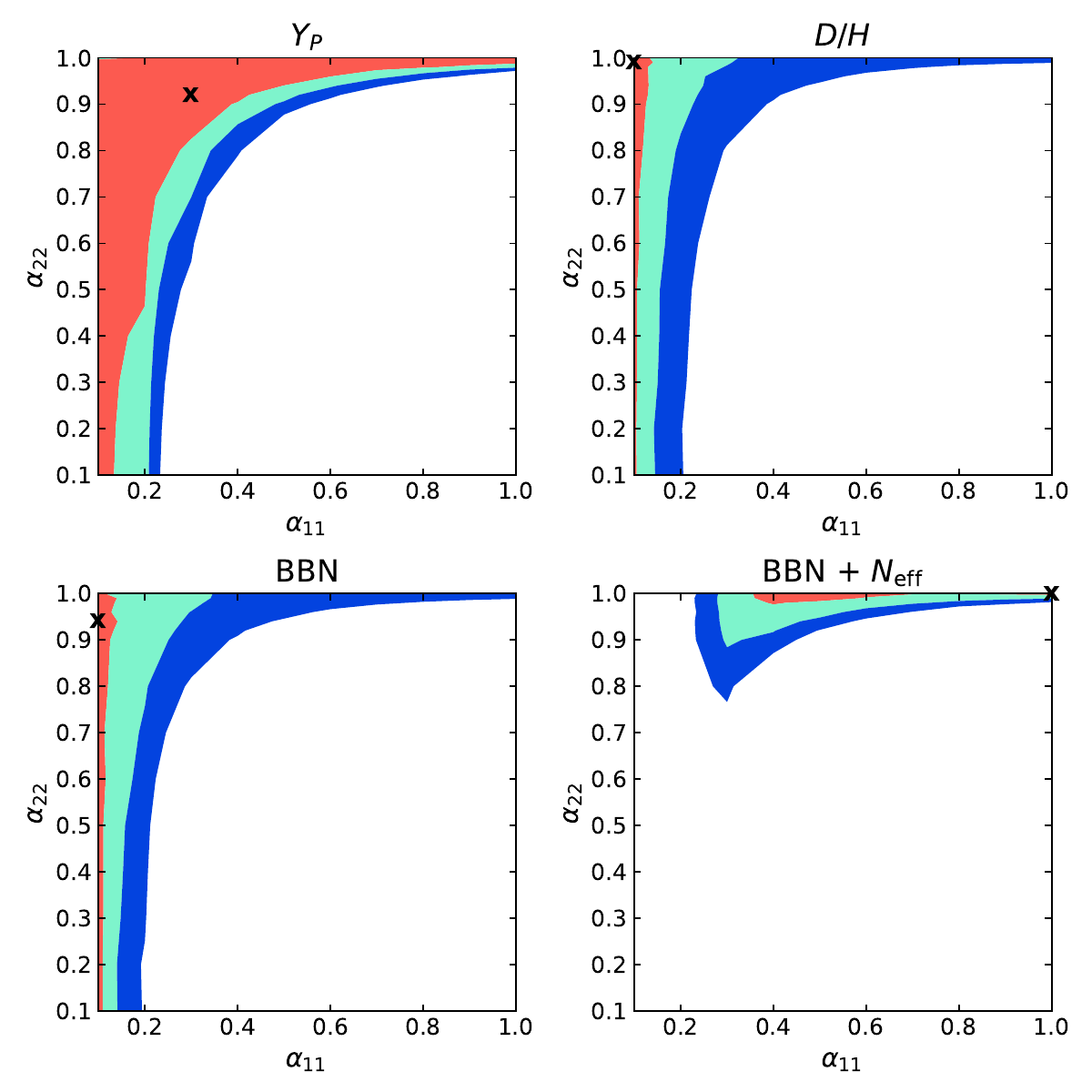}
    \caption{Same as Figure~\ref{fig:alpha_plane_NACRE} but for PRIMAT rates. The best-fit value for the combination of BBN with \Neff\ corresponds to the absence of non-unitarity.}
    \label{fig:alpha_plane_PRIMAT}
\end{figure}

Our approach is to include only the NACRE II rates because they are the ones that find a better degree of concordance with observations. We adopt a conservative stance and stick to the already very restrictive NACRE II results, but we still offer the PRIMAT results for completeness in Figure~\ref{fig:alpha_plane_PRIMAT}. Neutrino decoupling and the minimisation of $\Delta G_F$ remains unaltered by the choice of nuclear rates. Helium-4 results are fairly similar to the previously discussed. Unsurprisingly, PRIMAT rates yield much more constrained regions for deuterium, reflecting the tension already found in the SBBN. Since the PRIMAT prediction for deuterium is significantly lower than the observations, the $1\sigma$ region only appears in the region where $G^\beta_F$ does not change much and \Neff\ is also higher than the SM value (left side of the upper right panel). It is also the case that theoretical deuterium predictions where PRIMAT rates are more precise, being one order of magnitude smaller than with NACRE rates, which makes the $\chi^2$ more sensitive. In contrast to the previous analysis, the implementation of PRIMAT rates favours the SM over deviations from unitarity, as indicated by the best-fit of the BBN and \Neff\ combination. Furthermore, the constraints on the NU parameter space become stronger than before.

We conclude that the choice of the nuclear rates has profound implications and is a crucial aspect of the computation. Different rates either support or (mildly) penalise the SM when considering non-unitary three-neutrino mixing. This discrepancy is an instance of the potential challenges that may arise in BBN analysis until the settlement of the nuclear rates issue is resolved, either through the refinement of nuclear inputs or the addition of new deuterium observations.

The global picture is not expected to change even if further NU parameters are included ($\alpha_{3i}$), given that their impact on BBN yields is minimal. Neutrino decoupling may be affected but at the expense of departing more from the SM with a non-zero $\alpha_{3i}$~\cite{Gariazzo2022Non-unitarity}. We can thus safely conclude that the presented constraints can be considered as to be profiled over all NU parameters.

\section{Conclusions}
\label{sec:Conclusions}

Cosmology has entered its precision era, no longer aiming for order-of-magnitude estimations: cosmological observables, such as the primordial abundances of $^4He$ and $^2H$, are now measured with percent-level precision, while \Neff\ is expected to reach a similar precision in the near future. This exceptional level of precision has been achieved in the theoretical domain as well, thanks to the implementation of a thorough description of all known interactions and particles in the context of the evolution of the universe. In general terms, both theoretical and observational perspectives are consistent with each other, validating the well-established standard picture of both cosmology and particle physics. In light of this concordance, precision cosmology emerges as a powerful tool to probe BSM physics, testing its effect on the early stages of the universe and ruling out NP models that produce huge deviations from the standard value of the cosmological observables. 

In this work, we have explored two phenomenological BSM frameworks regarding potential mechanisms of neutrino mass generation: non-standard neutrino interactions and non-unitary three-neutrino mixing. We studied their impact on the early universe and cosmological observables, finding that neutrino NC-NSI with electrons mainly affect the thermodynamic background and thus \Neff, while neutrino CC-NSI with quarks modify the $n \leftrightarrow p$ conversion and thus the primordial abundances. In turn, non-unitarity involves both NC and CC interactions and, consequently, it affects both neutrino decoupling and BBN.

To quantify these effects, the BBN numerical code \prym\ has been adapted to study the change in BBN abundances in the presence of neutrino NC-NSI with electrons, neutrino CC-NSI with quarks and non-unitary three-neutrino mixing. The latter two have been studied for the first time in this context. 
In the case of neutrino NC-NSI with electrons, which primarily affect the thermodynamic background, BBN is considerably less constraining than \Neff. Alternatively, for neutrino CC-NSI with quarks and non-unitarity, BBN provides constraints that are comparable to or, in some cases, even more stringent than those from terrestrial experiments. In particular, we find that BBN sets competitive bounds to CC-NSI non-diagonal parameters $\varepsilon^{udV}_{\alpha \beta}$ and the NU diagonal parameter $\alpha_{22}$. When multiple NU parameters are considered, current observed nuclear abundances are found to be slightly more restrictive than future \Neff\ measurements, but both complement each other.

The combination of the results from both nuclear abundances and \Neff\ reveals that the non-unitarity parameter space is significantly constrained compared to what it would be if the two probes were considered separately. This is a perfect example of how neutrino decoupling and BBN can result in synergies that significantly enhance the constraining power of the early universe physics. Cosmological constraints, although still being indirect tests of NP, are always fundamental because they can complement terrestrial searches and serve as a consistency check. What is more, as demonstrated throughout this work, the highly precise cosmological observations are already providing competitive constraints.

Finally, we found that the choice of the nuclear rates can influence the interpretation of the non-unitarity results. When considering several NU parameters simultaneously, the results may either corroborate or slightly contradict the SM depending on the nuclear rates adopted. 
Although the tension remains statistically insignificant, if the precision is increased and it validates the PRIMAT rates, it could be the first signal of inconsistencies between BBN and the CMB. On the contrary, if the tension is resolved in favor of the concordance and the NACRE II rates, the cosmological constraints derived from BBN will be even more robust and stricter. Additional measurements of the nuclear rates of $D(d, n)^3He$ and $D(d, p)^3H$ are instrumental to resolve this puzzle~\cite{Pitrou2021ResolvingConclusions}. 

The future prospects for BBN observational determinations are less straightforward than the expected improvements in \Neff\ measurements. There is optimism for the upcoming next generation of $30-40$~m telescope facilities, that could enable a future detection of the $^3He/^4He$ ratio in extragalactic HII regions environments, that may reflect its primordial value~\cite{Cooke2015BBNandHe3}. They are also expected to increase by an order of magnitude the number of $D/H$ measurements, reducing its uncertainty and allowing for a robust evaluation of systematics~\cite{Cooke:2016rky}. Improvements in $^4He$ measurements are far less certain, but it is still conceivable to reach a subpercent precision~\cite{Grohs:2019Astro2020}. In the case of $^7Li$, further measurements combined with theoretical studies are required to elucidate the lithium problem~\cite{Sbordone2010Lithium}. This research highlights the importance of the established synergy between cosmological observations and particle physics experiments, exemplifying how precision measurements of the early universe can provide insights into the fundamental laws governing the cosmos.

\section*{Acknowledgements}
GB and ASV are supported by the Spanish grants  CIPROM/2021/054 (Generalitat Valenciana), PID2020-113775GB-I00 (MICIU/AEI/10.13039/501100011033), and by the European ITN project HIDDeN (H2020-MSCA-ITN-2019/860881-HIDDeN).
SG is supported by the Research grant TAsP (Theoretical Astroparticle Physics) funded by Istituto Nazionale di Fisica Nucleare (INFN).
ASV is also supported by the grant FPU23/01408, MICIU.

\bibliographystyle{JHEP}
\bibliography{biblio.bib}

\end{document}